\documentclass[aps,prl,preprint,tightenlines,superscriptaddress,showpacs,byrevtex,subfigure]{revtex4}

\usepackage{graphicx}
\usepackage{color}

\begin{document}
\begin{flushright}
KEK Preprint 2010-42\\
Belle Preprint 2010-24\\
NTLP  Preprint 2010-06\\
\end{flushright}
\title{ \quad\\[0.5cm] 
Search for {Lepton-Flavor-Violating} $\tau$ 
{Decays}
into a Lepton and a Vector Meson
}

\begin{abstract}

We search for {lepton-flavor-violating} 
{$\tau\to\ell V^0$ decays,
where $\ell$ is an electron or muon and $V^0$
is one of the vector mesons
$\rho^0,~\phi, ~\omega, ~K^{\ast 0}$ and $ \bar{K}^{\ast 0}$.
We use}
854 fb$^{-1}$ of data collected 
with the Belle detector at the 
KEKB asymmetric-energy $e^+e^-$ collider. 
No evidence for a signal {is} found
in 
{any decay mode,}
{and we} 
obtain 
{90\% confidence level upper limits}
 on the {individual branching 
fractions in the range $(1.2-8.4)\times 10^{-8}$.}

\end{abstract}

\affiliation{Budker Institute of Nuclear Physics, Novosibirsk, Russian Federation}
\affiliation{Faculty of Mathematics and Physics, Charles University, Prague, The Czech Republic}
\affiliation{Chiba University, Chiba, Japan}
\affiliation{University of Cincinnati, Cincinnati, OH, USA}
\affiliation{Justus-Liebig-Universit\"at Gie\ss{}en, Gie\ss{}en, Germany}
\affiliation{The Graduate University for Advanced Studies, Hayama, Japan}
\affiliation{Hanyang University, Seoul, South Korea}
\affiliation{University of Hawaii, Honolulu, HI, USA}
\affiliation{High Energy Accelerator Research Organization (KEK), Tsukuba, Japan}
\affiliation{Hiroshima Institute of Technology, Hiroshima, Japan}
\affiliation{University of Illinois at Urbana-Champaign, Urbana, IL, USA}
\affiliation{Institute of High Energy Physics, Chinese Academy of Sciences, Beijing, PR China}
\affiliation{Institute for High Energy Physics, Protvino, Russian Federation}
\affiliation{Institute of High Energy Physics, Vienna, Austria}
\affiliation{Institute for Theoretical and Experimental Physics, Moscow, Russian Federation}
\affiliation{J. Stefan Institute, Ljubljana, Slovenia}
\affiliation{Kanagawa University, Yokohama, Japan}
\affiliation{Institut f\"ur Experimentelle Kernphysik, Karlsruher Institut f\"ur Technologie, Karlsruhe, Germany}
\affiliation{Korea Institute of Science and Technology Information, Daejeon, South Korea}
\affiliation{Korea University, Seoul, South Korea}
\affiliation{Kyungpook National University, Taegu, South Korea}
\affiliation{\'Ecole Polytechnique F\'ed\'erale de Lausanne, EPFL, Lausanne, Switzerland}
\affiliation{Faculty of Mathematics and Physics, University of Ljubljana, Ljubljana, Slovenia}
\affiliation{University of Maribor, Maribor, Slovenia}
\affiliation{Max-Planck-Institut f\"ur Physik, M\"unchen, Germany}
\affiliation{University of Melbourne, Victoria, Australia}
\affiliation{Nagoya University, Nagoya, Japan}
\affiliation{Nara Women's University, Nara, Japan}
\affiliation{National Central University, Chung-li, Taiwan}
\affiliation{National United University, Miao Li, Taiwan}
\affiliation{Department of Physics, National Taiwan University, Taipei, Taiwan}
\affiliation{H. Niewodniczanski Institute of Nuclear Physics, Krakow, Poland}
\affiliation{Nippon Dental University, Niigata, Japan}
\affiliation{Niigata University, Niigata, Japan}
\affiliation{University of Nova Gorica, Nova Gorica, Slovenia}
\affiliation{Novosibirsk State University, Novosibirsk, Russian Federation}
\affiliation{Osaka City University, Osaka, Japan}
\affiliation{Panjab University, Chandigarh, India}
\affiliation{University of Science and Technology of China, Hefei, PR China}
\affiliation{Seoul National University, Seoul, South Korea}
\affiliation{Sungkyunkwan University, Suwon, South Korea}
\affiliation{School of Physics, University of Sydney, NSW 2006, Australia}
\affiliation{Tata Institute of Fundamental Research, Mumbai, India}
\affiliation{Excellence Cluster Universe, Technische Universit\"at M\"unchen, Garching, Germany}
\affiliation{Toho University, Funabashi, Japan}
\affiliation{Tohoku Gakuin University, Tagajo, Japan}
\affiliation{Tohoku University, Sendai, Japan}
\affiliation{Department of Physics, University of Tokyo, Tokyo, Japan}
\affiliation{Tokyo Metropolitan University, Tokyo, Japan}
\affiliation{Tokyo University of Agriculture and Technology, Tokyo, Japan}
\affiliation{CNP, Virginia Polytechnic Institute and State University, Blacksburg, VA, USA}
\affiliation{Yonsei University, Seoul, South Korea}
\author{Y.~Miyazaki} % Nagoya 
\affiliation{Nagoya University, Nagoya, Japan}
\author{H.~Aihara} % Tokyo 
\affiliation{Department of Physics, University of Tokyo, Tokyo, Japan}
\author{K.~Arinstein} % BINP  Novosibirsk 
\affiliation{Budker Institute of Nuclear Physics, Novosibirsk, Russian Federation}
\affiliation{Novosibirsk State University, Novosibirsk, Russian Federation}
\author{V.~Aulchenko} % BINP  Novosibirsk 
\affiliation{Budker Institute of Nuclear Physics, Novosibirsk, Russian Federation}
\affiliation{Novosibirsk State University, Novosibirsk, Russian Federation}
\author{A.~M.~Bakich} % Sydney 
\affiliation{School of Physics, University of Sydney, NSW 2006, Australia}
\author{V.~Balagura} % ITEP 
\affiliation{Institute for Theoretical and Experimental Physics, Moscow, Russian Federation}
\author{E.~Barberio} % Melbourne 
\affiliation{University of Melbourne, Victoria, Australia}
\author{A.~Bay} % Lausanne 
\affiliation{\'Ecole Polytechnique F\'ed\'erale de Lausanne, EPFL, Lausanne, Switzerland}
\author{K.~Belous} % Protvino 
\affiliation{Institute for High Energy Physics, Protvino, Russian Federation}
\author{V.~Bhardwaj} % Panjab 
\affiliation{Panjab University, Chandigarh, India}
% ...
\collaboration{The Belle Collaboration}
\noaffiliation

\pacs{11.30.Fs; 13.35.Dx; 14.60.Fg}
%Belle Collaboration}
\maketitle
 \section{Introduction}

{Lepton flavor violation (LFV)
in charged lepton decays is forbidden 
{in the Standard Model (SM)} 
%and highly suppressed
%if neutrino mixing is included.
{and remains highly suppressed 
even if the SM is modified to include neutrino mixing.}
However, 
extensions of the SM,
such as supersymmetry, leptoquark and many other 
models~\cite{rpv,cite:amon,cite:six_fremionic,cite:susy1,cite:susy2,Benbrik:2008ik,Li:2009yr}
predict LFV with branching fractions
as high as {$10^{-8}$,}
{which could already be
accessible
in
current
$B$-factory experiments.}
We {search} for 
{$\tau^-\to\ell^- V^0$ decays{\footnotemark[2]},
where $\ell$ is an electron or muon and $V^0$
is one of the vector mesons
$\rho^0,~\phi, ~\omega, ~K^{\ast 0}$ {or} $ \bar{K}^{\ast 0}$.}
{These results are based on the entire 854 fb$^{-1}$ data sample}
%with a total data sample of 854 fb$^{-1}$ 
collected 
{at center-of-mass (CM) energies near the $\Upsilon(4S)$,
near the $\Upsilon(5S)$,}
and off resonance with the Belle detector~\cite{Belle} at the KEKB 
asymmetric-energy   $e^+e^-$ collider~\cite{kekb}.
Previously, we
obtained
90\% confidence level (C.L.) upper limits
{on} branching 
{fractions of these decays using}
543 fb${}^{-1}$ of data;
the results were
in the range (5.8$-$18)~$\times~10^{-8}$~\cite{lv0_belle}.
{}The results described here use additional data
and an improved event selection procedure,
which is optimized mode-by-mode.}
The BaBar collaboration
has
also
published {90\% C.L.
upper limits}
in the range (2.6$-$19)~$\times~10^{-8}$
using 451 fb${}^{-1}$ of data~\cite{lv0_babar}
{for all $\tau^-\to\ell^- V^0$ decays 
{except for} 
{$\tau^-\to\ell^-\omega$} 
for which 384 fb${}^{-1}$ of data were used~\cite{lomega_babar}.}

\footnotetext[2]{{Throughout this paper,
charge-conjugate modes are {implied} unless stated  otherwise.}}

The Belle detector is a large-solid-angle magnetic spectrometer that
consists of a silicon vertex detector (SVD), 
a 50-layer central drift chamber (CDC), 
an array of aerogel threshold 
{{C}herenkov} counters (ACC), a barrel-like arrangement of 
time-of-flight scintillation counters (TOF), and an electromagnetic calorimeter 
comprised of  
CsI(Tl) {crystals (ECL), all located} inside
a superconducting solenoid coil
that provides a 1.5~T magnetic field.  
An iron flux-return located outside of the coil is instrumented to detect 
$K_{\rm{L}}^0$ mesons 
and to identify muons (KLM).  
The detector is described in detail elsewhere~\cite{Belle}.

Particle identification
is very important
{for this measurement.}
{We use 
{particle} identification likelihood variables} 
based on
the ratio of the energy
deposited in the
ECL to the momentum measured in the SVD and CDC,
shower shape in the ECL,
the {particle's} range in the KLM,
hit information from the ACC,
{$dE/dx$ {measured} in the CDC,}
and {the {particle's time of flight}}.
To distinguish hadron species,
we use likelihood ratios,
${\cal{P}}(i/j) = {\cal{L}}_i/({\cal{L}}_i + {\cal{L}}_{j})$,
where ${\cal{L}}_{i}$ (${\cal{L}}_{j}$)
is the likelihood {of} the {observed} 
detector response
{for} {a} track with flavor $i$ ($j$).
{{For lepton identification,
we {form} likelihood ratios ${\cal P}(e)$~\cite{EID}
and ${\cal P}({\mu})$~\cite{MUID}}
{using}
the responses of the appropriate subdetectors.

{We use {Monte Carlo} (MC) samples}
to estimate the signal efficiency and 
optimize the event selection.
{Signal} and background events from generic $\tau^+\tau^-$ decays are 
generated by KKMC/TAUOLA~\cite{KKMC}. 
{For signal,} 
we generate {$e^+e^-\to\tau^+\tau^-$} {events,} 
{in which} 
{one} $\tau$ {is forced to decay}  
into  {a lepton} 
and a vector {meson}
{using {a two-body {phase space} model,}}
{while} 
the other $\tau$ decays generically.
{Background events
$B\bar{B}$, continuum
$e^+e^-\to q\bar{q}$ ($q=u,d,s,c$), Bhabha,} 
and two-photon processes are generated by 
EvtGen~\cite{evtgen},
BHLUMI~\cite{BHLUMI}, 
and
{AAFH~\cite{AAFH}}, respectively. 
{In what follows,}
all kinematic variables are calculated in the laboratory frame
unless otherwise specified.
In particular,
variables
calculated in the $e^+e^-$ {CM frame}
are indicated by the superscript ``CM''.

\section{Event Selection}

%
% Event topology 
%

{We search for {$e^+e^-\to\tau^+\tau^-$} 
events in which one $\tau$ 
(the {signal $\tau$}) decays into a lepton
and
a vector meson  ($V^0 =~\rho^0,~\phi,~\omega,~K^{\ast 0}$ and 
$\bar{K}^{\ast 0}$)
and the other $\tau$~(the {tag $\tau$}) 
{results in}
{a single} charged track, any number of additional 
photons, and neutrinos.
We reconstruct 
$\rho^0$ from $\pi^+\pi^-$, $\phi$ from $K^+ K^-$,
$\omega$ from $\pi^+\pi^-\pi^0$,
$K^{*0}$ from $K^+ \pi^-$ and $\bar{K}^{*0}$ from $K^- \pi^+$.
{To improve 
%on our previous work
{sensitivity compared to our previous work}~\cite{lv0_belle},
the event selection is optimized {mode-by-mode},}
since the {backgrounds} are mode dependent.}

{For each candidate event we calculate}
the $\ell V^0$ invariant
mass~($M_{\rm {\ell V^0}}$) {and} 
the difference of {the $\ell V^0$} energy from the 
beam energy in the CM {frame}~($\Delta E$).
{In the 
two-dimensional distribution of $M_{\rm {\ell V^0}}$ versus $\Delta E$,}
{signal events} 
should have $M_{\rm {\ell V^0}}$
close to the $\tau$-lepton mass ($m_{\tau}$) and
$\Delta E$ close to zero.

%
% track
%

{We require that events contain}
four charged tracks and any number of photons within the fiducial volume
 defined by $-0.866 < \cos\theta < 0.956$,
{where $\theta$ is 
the polar angle {relative} to 
the direction opposite to 
that of 
the {incident} $e^+$ beam in 
{the} laboratory frame.}
The transverse momentum ($p_t$) of each charged track
and {the} energy of each photon ($E_{\gamma}$) 
are 
{required to satisfy} $p_t> $ 0.1 GeV/$c$ and $E_{\gamma}>0.1$ GeV,
respectively.
{For each charged track, 
the distance of the closest point with 
respect to the interaction point 
is required to be 
less than 0.5 cm in the transverse direction 
and less than 
3.0 cm in the longitudinal direction.}

%
% lepton ID and gamma veto in signal side
%

{Using the plane perpendicular to the CM
thrust axis~\cite{thrust},
which is calculated from 
the observed tracks and photon candidates,
we separate the particles in an event
into two hemispheres.
These  are referred to as the signal and 
tag sides. }
{The tag side is required to contain one charged track
while the signal side is required to 
{contain three charged tracks.}}
We require one charged track on the signal side 
{to be} 
identified as a lepton.
{Electron candidates are required 
to have ${\cal P}(e) > 0.9$ and
{momenta} $p > 0.6$ GeV/$c$
while muon candidates should have 
${\cal P}(\mu) > 0.95$ and $p >1.0 $ GeV/$c$.}
%{The electron and muon {identification} criteria are}
%${\cal P}(e) > 0.9$ with 
%{momentum} $p > 0.6$ GeV/$c$
%and 
%${\cal P}(\mu) > 0.95$ with $p >1.0 $ GeV/$c$,
%respectively.
In order to take into account the emission
of  bremsstrahlung photons from {electrons,}
the momentum of the
electron track
is reconstructed 
{{by adding to it the} 
momentum of every photon
within}
0.05 radians 
{of}
{the track {direction}.}
{With these requirements, the} electron (muon) identification
{efficiency
is} 91\% (85\%)
while
{the probability to misidentify {a} pion
as
{an} {electron} ({a} muon)}
is below 0.5\% (2\%).

\begin{figure}
\begin{center}
       \resizebox{0.35\textwidth}{0.35\textwidth}{\includegraphics
        {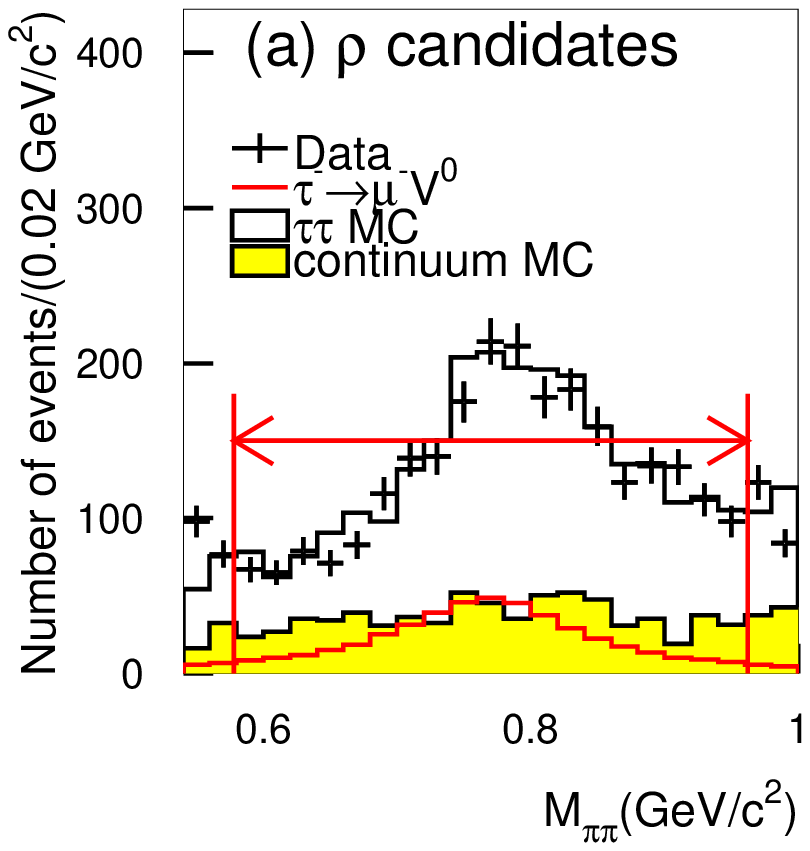}}
 \hspace*{-1.cm}
       \resizebox{0.35\textwidth}{0.35\textwidth}{\includegraphics
        {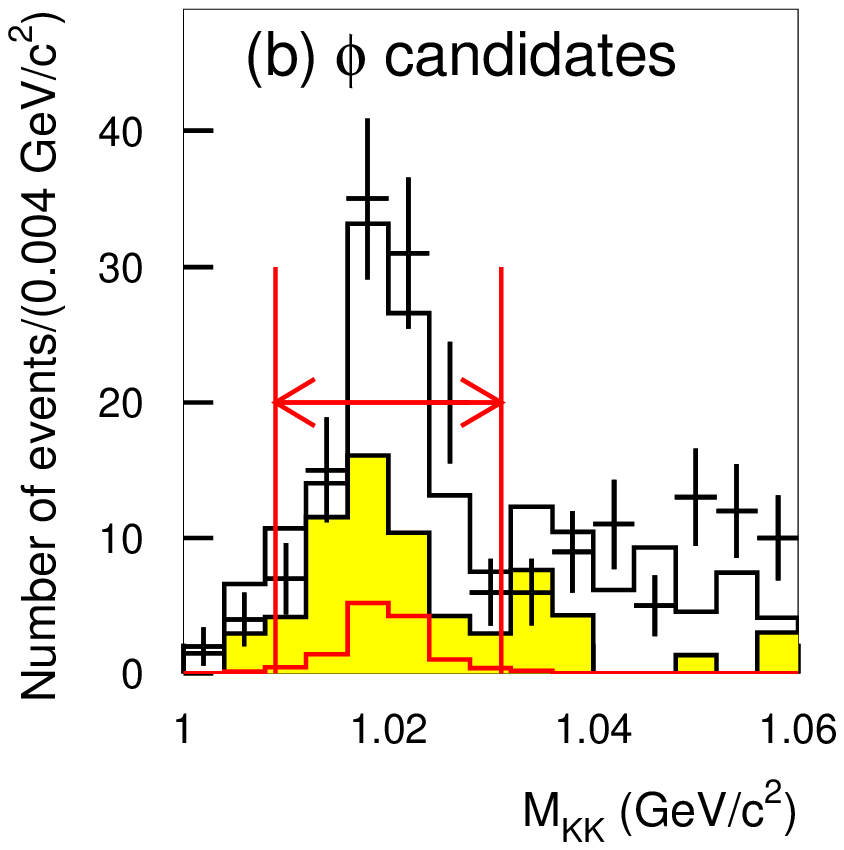}}
 \hspace*{-1.cm}
       \resizebox{0.35\textwidth}{0.35\textwidth}{\includegraphics
        {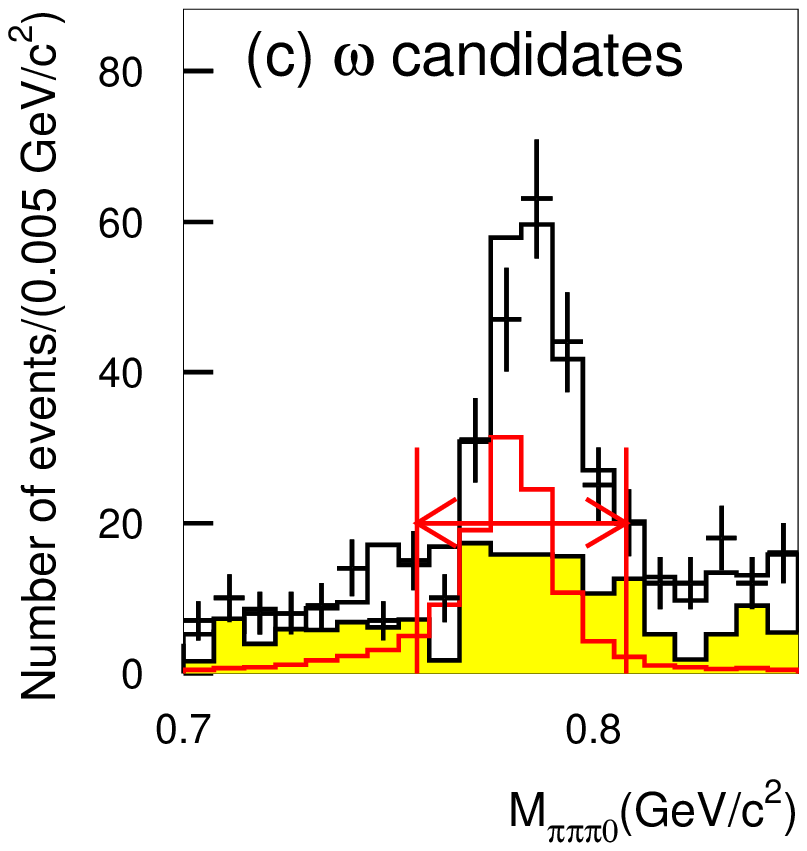}}\\
 \hspace*{-5.3cm}
       \resizebox{0.35\textwidth}{0.35\textwidth}{\includegraphics
 {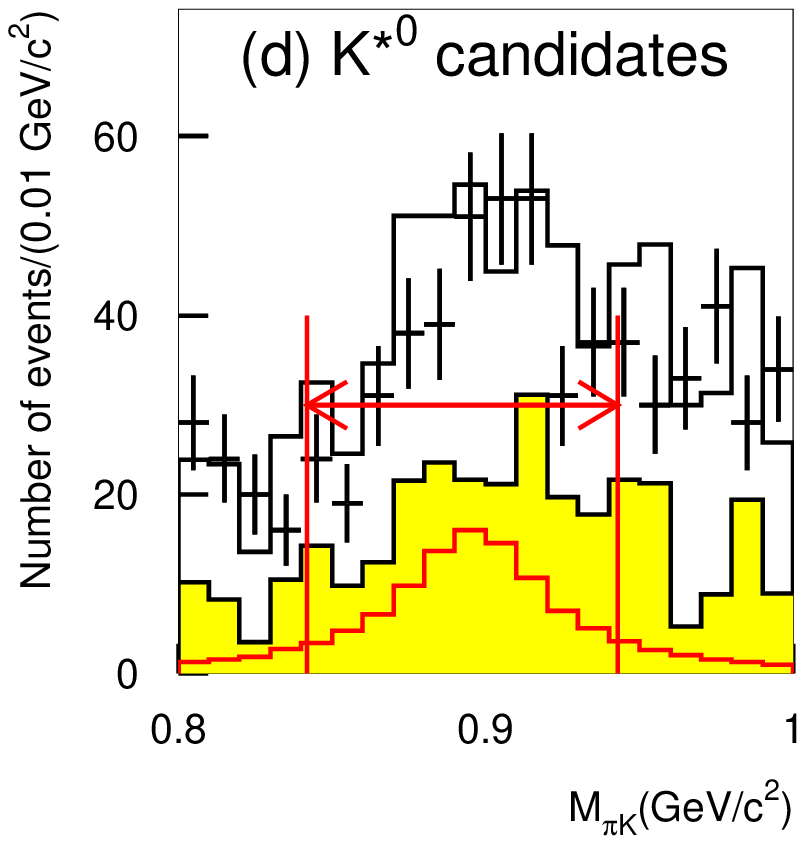}}
 \hspace*{-1.cm}
       \resizebox{0.35\textwidth}{0.35\textwidth}{\includegraphics
 {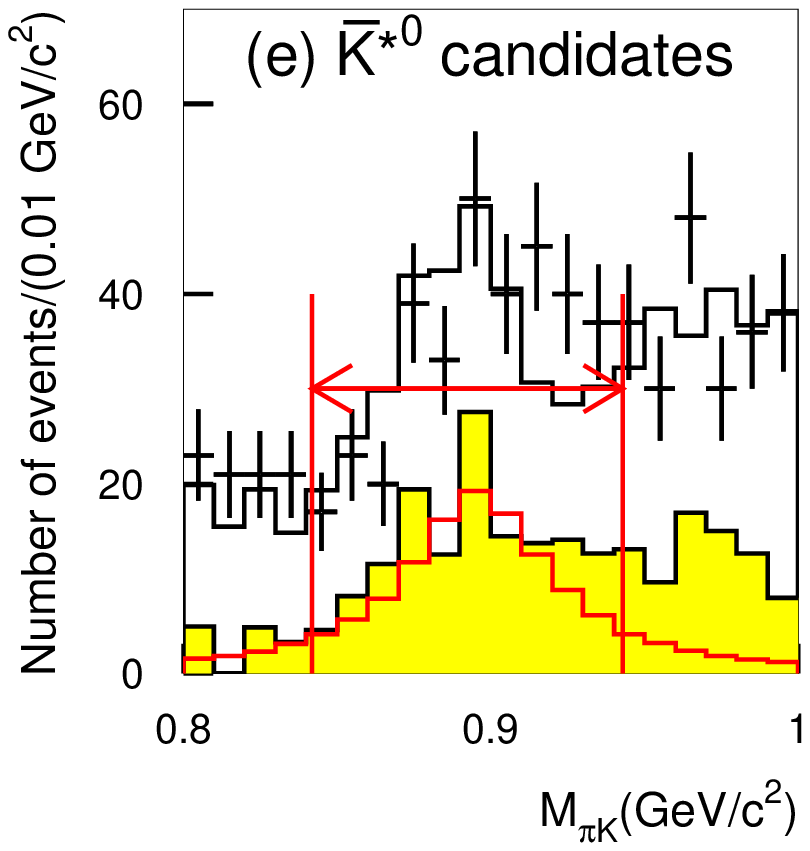}}\\
\caption{Vector meson invariant {mass} 
distributions  for the $\tau\to\mu V^0$ modes
after {the}
baseline selection 
({before} vector meson {mass requirements})
{in the region} $1.5$ GeV/$c^2$ $ <M_{\ell V^0} < 2.0$ GeV/$c^2$ and 
$-0.5$ GeV $< \Delta E < 0.5$ GeV:
(a), (b), (c), (d) and (e) show
the $\tau\to\mu\rho^0$, $\mu\phi$, $\mu\omega$, $\mu {K}^{\ast0}$ and 
$\mu \bar{K}^{\ast0}$ {modes,} respectively.
{The signal MC ($\tau\to\mu V^0$)
distributions {are} {normalized} arbitrarily 
while the background MC
distributions are normalized to the data luminosity.}
{Selected {mass} regions are indicated
by 
arrows. }}
\label{fig:v0mass}
\end{center}
\end{figure}

\begin{table}
\caption{
{Allowed ranges}  for  the mass 
%region for {a} 
{of}
$V^0$ {candidates} 
and 
{for}
the cosine of the  angle 
between {the} missing particle and the tag-side track
in the CM system ($\cos \theta^{\rm CM}_{\rm tag-miss}$).
}
\label{v0_masscut}
\begin{tabular}{|c|c|c|}\hline\hline
$V^0$ & Invariant mass (GeV/$c^2$)& $\cos \theta^{\rm CM}_{\rm tag-miss}$ for $\tau\to\mu V^0$ ($e V^0$)
\\ \hline
$\rho^0$ & $0.587 < M_{\pi\pi} < 0.962$ &  [0.0, 0.85] ([0.0, 0.96])\\
$\phi$ & $1.009 < M_{KK} < 1.031$ &  [0.0, 0.88] ([0.0, 0.97])\\
$\omega$ & $0.757 < M_{\pi\pi\pi} < 0.808$ &  [0.0, 0.88] ([0.0, 0.97])\\
$K^{0\ast} (\bar{K}^{0\ast})$ & $0.842 < M_{K\pi} < 0.956$ &  [0.0, 0.87] ([0.0, 0.96])\\
\hline\hline
\end{tabular}
\end{table}

%
% K/pi selection
%

Charged kaons are identified by
{the} 
condition ${\cal{P}}(K/\pi) > 0.8$ for 
the $\tau\to\ell K^{\ast 0}$ and $\ell \bar{K}^{\ast 0}$   modes
($>0.6$ for the $\ell\phi$ modes).
Charged pions 
%are identified by {the} condition 
{must satisfy the requirement} 
${\cal{P}}(K/\pi) < 0.6$ for 
the $\tau\to\ell\rho^0$, $\ell K^{\ast 0}$ and $\ell \bar{K}^{\ast 0}$  modes.
We {do not}  require  charged pion identification for the $\tau\to\ell\omega$ modes.
The kaon (pion) identification
{efficiency is} 80\% (88\%)
while
{the probability to misidentify {a} pion
(kaon)
as
{a} {kaon} ({a} pion)}
is below 10\% (12\%).
{In order to 
{reduce the fake vector meson background}
from {photon} conversions
(i.e., $\gamma \rightarrow e^+e^-$),}
we require that {the} 
{two charged tracks from the meson candidate}
have ${\cal{P}}(e) <0.1$.
Furthermore, we require 
${\cal{P}(\mu)} <0.1$
to suppress 
$e^+e^- \to e^+e^-\mu^+\mu^-$ {two-photon background.}
For the $\tau\to\ell\omega$ modes, we select $\pi^0(\to\gamma\gamma)$ 
candidates on the signal side {by}
{requiring} $E_\gamma > 0.1$ GeV and
$p_{\pi^0} > 0.4$ GeV/$c$, {which suppresses}  
{incorrectly reconstructed} $\pi^0$ candidates from
initial-state radiation (ISR) {and  beam background.}  
The $\pi^0$ mass window is 
0.12 GeV/${c}^2$ $< M_{\gamma\gamma} < $ 0.15 GeV/${c}^2$.

%
% missing mass & p_miss
%

To ensure that the missing particles are neutrinos rather
than photons or charged particles that pass  outside the detector acceptance,
we impose requirements on the missing 
momentum $\vec{p}_{\rm miss}$,
which 
is
calculated by subtracting the
vector sum of the momenta of all tracks and photons
from the sum of the $e^+$ and $e^-$ beam momenta.
We require that {$|p^{\rm{t}}_{\rm miss}|$,}
the magnitude of {the} transverse  {component of}
{$\vec{p}_{\rm miss}$}, 
be  greater than 0.5 GeV/$c$ (0.7 GeV/$c$) 
for the $\tau\to\mu V^0$ ($eV^0$ {except} the $e\rho^0$) modes,
and {that} 
{{$\vec{{p}}_{\rm miss}$} point} 
into 
{the {fiducial} volume} of the
detector.
For the $\tau\to e\rho^0$ mode only,
we apply {{the} {more restrictive} 
selection requirement 
{$|{p}^{\rm t}_{\rm miss}|$}  $>$ 1.5 GeV/$c$.}
Furthermore,
we reject events if the  
direction of the missing momentum 
{traverses the
gap between the barrel and the endcap} 
since an undetected photon 
{may be incorrectly attributed to}
missing particles.

To suppress  $B\bar{B}$ and $q\bar{q}$ background,
we require that 
the number of photons on the tag side, $n_{\gamma}^{\rm{TAG}}$,
{satisfy} 
$n_{\gamma}^{\rm{TAG}} \le 2$  $(\le 1)$
for hadronic (leptonic) {tags.}
{For all modes except $\pi^0\to\gamma\gamma$,
we also require at most one photon on the signal side.}

We {select}  
{vector mesons}
whose invariant mass  
{satisfies} the  {requirements}
shown in Table~\ref{v0_masscut} and Fig.~\ref{fig:v0mass}.
{{MC simulation {predicts that}
for the $\tau\to\mu V^0$ modes
the dominant background 
comes from continuum and generic $\tau^+\tau^-$events,
{whereas} for the $\tau\to eV^0$ modes 
it 
{originates} 
from 
inelastic $V^0$-photoproduction ($e^+e^-\to e^+e^-V^0$)
and two-photon processes.}

Since neutrinos are 
emitted on the tag side only,
the direction of
$\vec{p}_{\rm miss}$
should lie within the tag side of the event.
The cosine of the
opening angle between
$\vec{p}_{\rm miss}$
and the charged track on the tag side 
{in the CM system,}
$\cos \theta^{\mbox{\rm \tiny CM}}_{\rm tag-miss}$, 
{is required to be greater than zero.}
{If {the} track on the tag side is {a} {hadron},
then {for the $\tau\to \mu V^0$ modes} we} also require 
$\cos \theta^{\mbox{\rm \tiny CM}}_{\rm tag-miss}<(0.85-0.88)$. 
{This $\cos \theta^{\rm CM}_{\rm tag-miss}$ requirement} 
reduces continuum background 
{with}
missing 
{energy due to}
neutrons {or $K_L$'s}
since the {masses} 
of the neutron and $K_L$ are {higher} 
than that of {the} 
neutrino.
We also require that $\cos \theta^{\mbox{\rm \tiny CM}}_{\rm tag-miss}<(0.96-0.97)$ 
for the $\tau\to e V^0$ modes.
{This requirement}
can reduce Bhabha, 
inelastic $V^0$-photoproduction,
and two-photon background,
since radiated photons
from the tag-side track
result in missing momentum
if they overlap with
the ECL
clusters {associated with}
the tag-side track}~\cite{cite:tau_egamma}.}
{The $\cos \theta^{\mbox{\rm \tiny CM}}$ requirements for all modes 
{are given} in Table~\ref{v0_masscut}.}

\begin{figure}
\begin{center}
       \resizebox{0.35\textwidth}{0.3\textwidth}{\includegraphics
        {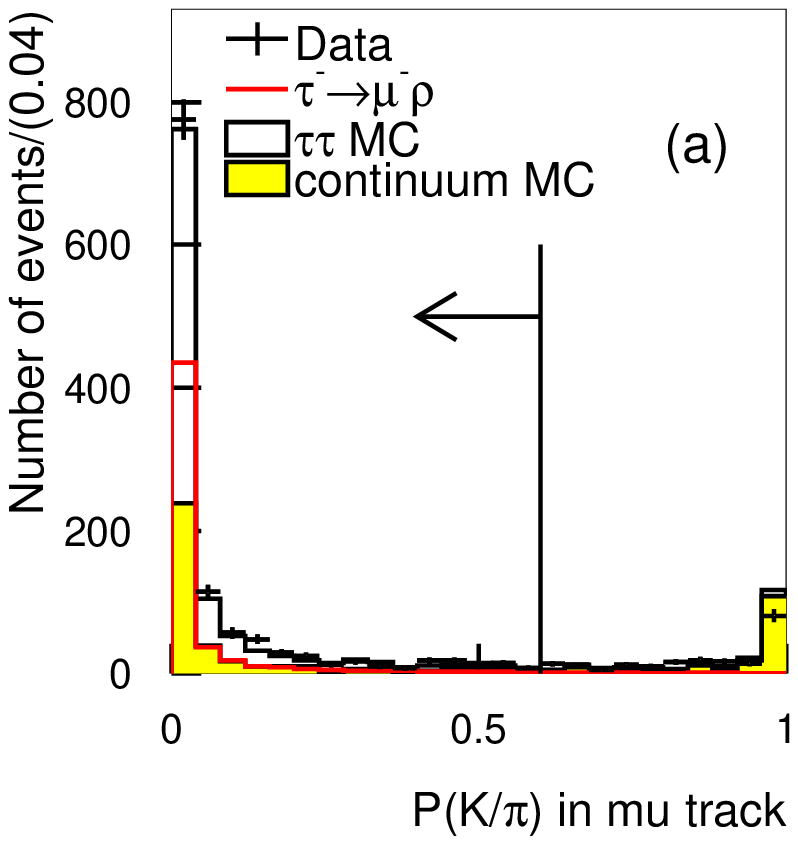}}
\hspace*{-1.cm}
       \resizebox{0.35\textwidth}{0.3\textwidth}{\includegraphics
        {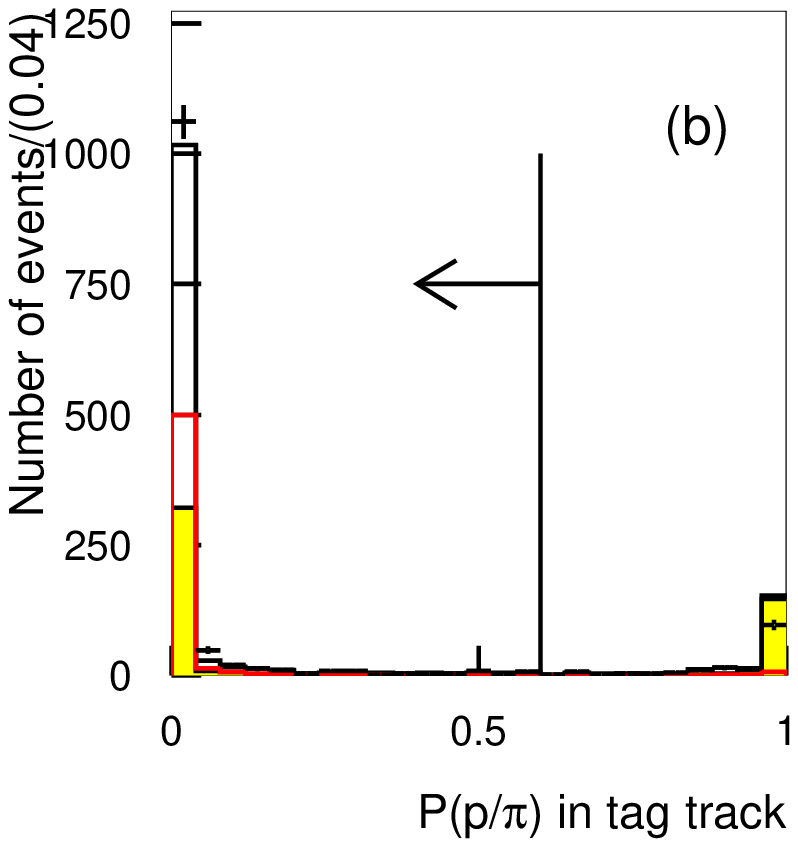}}
\hspace*{-1.cm}
       \resizebox{0.35\textwidth}{0.3\textwidth}{\includegraphics
        {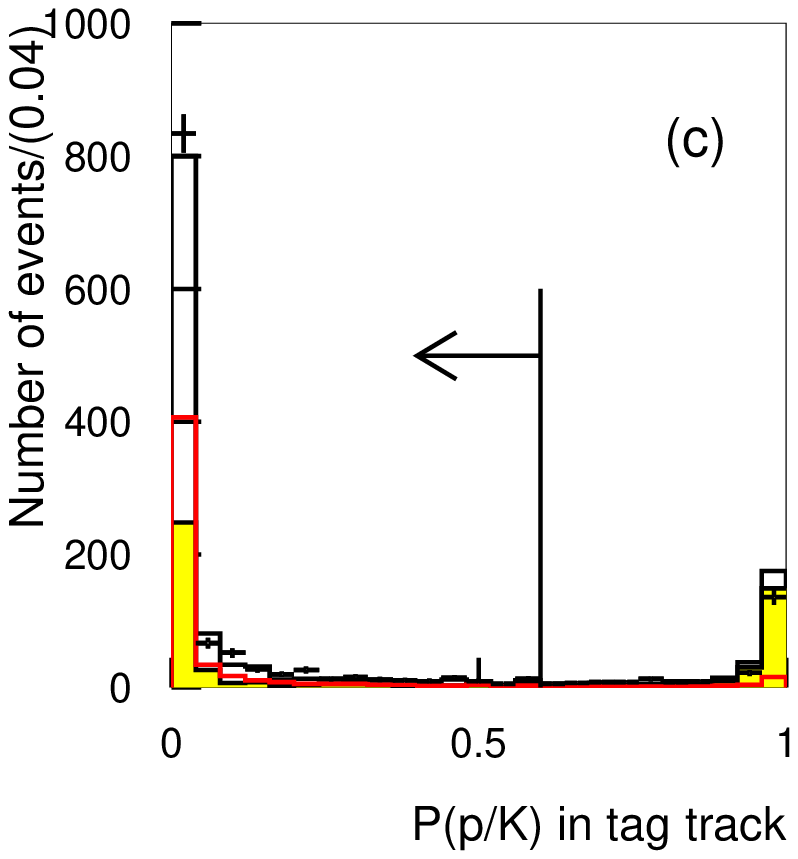}}
 \vspace*{-0.5cm}
\caption{
(a) ${\cal{P}}(K/\pi)$ for muon {tracks,} 
(b) ${\cal{P}}(p/\pi)$ and (c) ${\cal{P}}(p/K)$ for  
%tag {tracks} with a 
hadronic {tags, for $\mu\rho^0$ candidate events satisfying}
$1.5$ GeV/$c^2$ $ <M_{\ell V^0} < 2.0$ GeV/$c^2$ and 
$-0.5$ GeV $< \Delta E < 0.5$ GeV. 
{{Signal} MC ($\tau\to\mu\rho^0$)
distributions {are} {normalized} arbitrarily 
while the background MC
 distributions are normalized to the data luminosity.}
 {Selected regions are indicated
 by {arrows}.
}}
\label{fig:pid}
\end{center}
\end{figure}

For the $\tau\to\mu V^0$ modes,
a {muon} 
{can originate} from a kaon decaying in the CDC ($K^\pm\to\mu^{\pm}\nu_\mu$). 
Therefore, 
we apply a kaon veto, ${\cal{P}}(K/\pi)<0.6$, 
for 
muon candidate tracks if {{the} tag side track} is {a}  hadron
(see Fig.~\ref{fig:pid}~(a)).
{Another important continuum background }
{predicted by MC}
{is} from dibaryon production with a proton on the tag side.
Therefore, 
we apply a proton veto, ${\cal{P}}(p/\pi)<0.6$ and ${\cal{P}}(p/K)<0.6$, 
{as shown in  Figs.~\ref{fig:pid} (b) and (c).}
To reject $q\bar{q}$ continuum background,
we require that the magnitude of thrust ($T$) be greater than 0.90.
The reconstructed 
{mass
of {the} 
charged track 
and photons}
on the tag side, 
$m_{\rm tag}$, 
is required to be less than the {$\tau$ lepton} mass (1.777 GeV/$c^2$).
To reduce $e^+e^-\to\mu^+\mu^-$ background for the $\tau\to\mu\rho^0$ mode only,
we also require that the momentum of the track on the tag side
in the CM system be less than 4.0 GeV/c 
{if the track is a muon candidate.}

\begin{figure}
\begin{center}
       \resizebox{0.32\textwidth}{0.3\textwidth}{\includegraphics
        {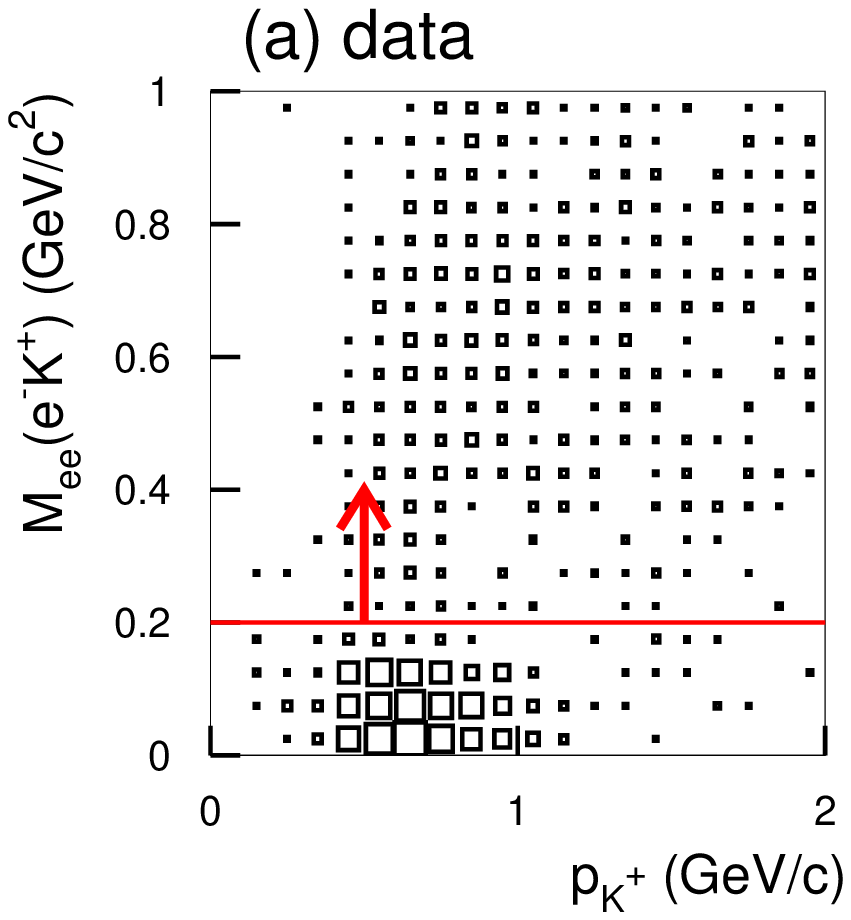}}
       \resizebox{0.32\textwidth}{0.3\textwidth}{\includegraphics
        {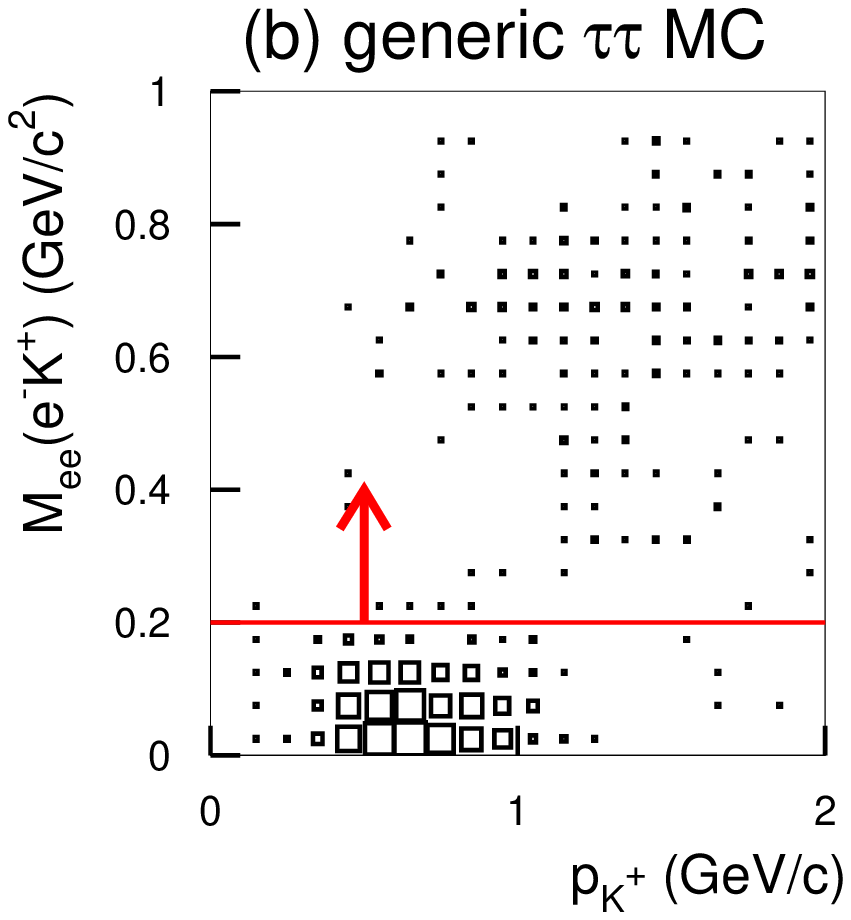}}
       \resizebox{0.32\textwidth}{0.3\textwidth}{\includegraphics
        {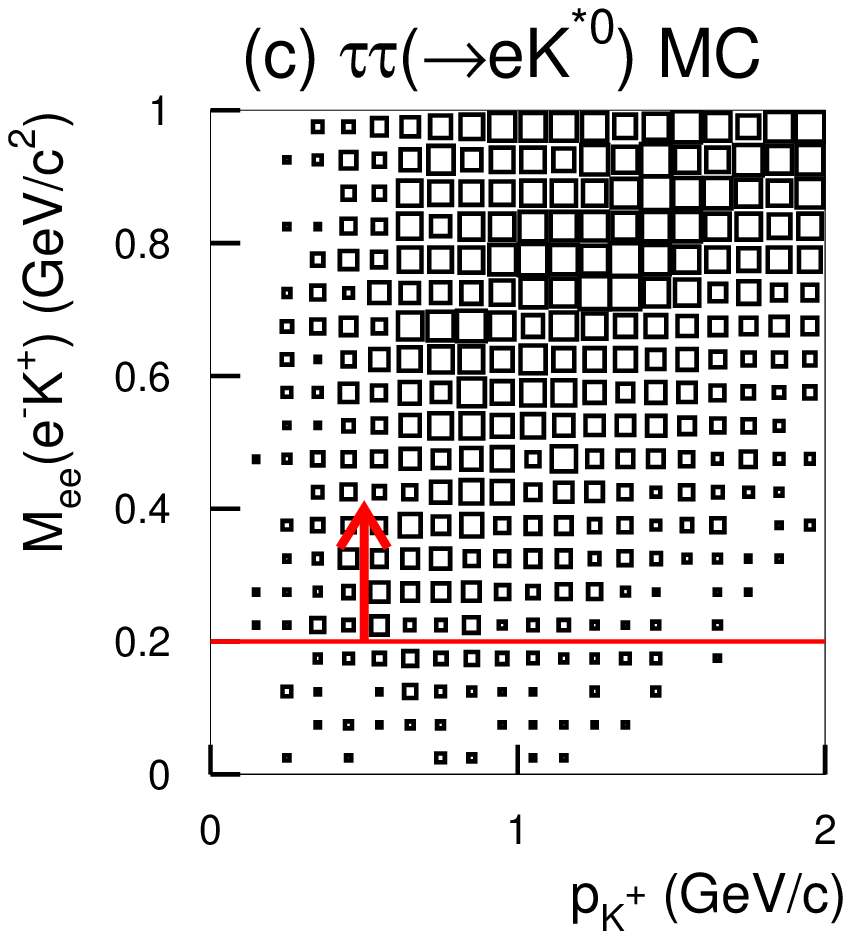}}
 \vspace*{-0.5cm}
\caption{Scatter-plot {of}
the momentum of kaon candidates from $K^{\ast0}$ candidates, $p_{K^+}$, 
{versus} 
the invariant mass of electron and kaon candidates from $\tau\to eK^{\ast0}(\to K^+\pi^-)$ 
reconstructed  by assigning the electron mass to the kaon tracks, 
$M_{ee}$($e^-K^+$),
for the $e K^{\ast0}$ mode:
(a), (b) and (c)  show 
the distributions for data, generic $\tau\tau$ {MC} 
and  $\tau\tau (\to e {K}^{\ast0})$ MC, respectively.
Selected regions are indicated by {red lines and arrows}.}
\label{fig:kstar_conv}
\end{center}
\end{figure}

For  the $\tau\to eV^0$ modes, 
photon conversions
can
result in large backgrounds
when {an $e^+e^-$ pair}
is {incorrectly {reconstructed} as a} 
fake 
vector meson.
{{Fake vector mesons} {can also originate from
one of {the} {electrons} {in} a photon conversion} and a hadron.} 
{As seen from generic $\tau^+\tau^-$ MC, 
such  {a}
background  comes from  $\tau^-\to h^-\pi^0(\to\gamma\gamma)\nu_{\tau}$ decays.}
When a $\gamma$ from {the}
$\pi^0$-decay converts and one of the {conversion}
electrons 
{traverses the ECL gap, then it 
may incorrectly be classified as} 
a hadron because 
the electron veto does not work. 
In the momentum range between 0.5 GeV/$c$ and 1.0 GeV/$c$ such an electron can 
be {misidentified} as a kaon because
{the} $dE/dx$ {distributions of
kaons and electrons overlap.} 
In {other momentum ranges the}
electron can 
be misidentified as a pion. 
If we assign an electron mass
to such a fake hadron, then 
the invariant mass $M_{ee}(e^-h^+)$ 
will be small, {as expected} 
for conversions. 
Therefore we require 
$M_{ee}(e^-h^+) (M_{ee}(h^-h^+)) >  0.2$ GeV/$c^2$ for $e^-h^+$ ($h^-h^+$) 
candidates 
for {the} $e\rho^0$, $eK^{\ast0}$ and $e\bar{K}^{\ast0}$ modes.
{Background from}
such $e^-K^+$ events  in the $e K^{\ast 0}$ mode
is shown in Fig.~\ref{fig:kstar_conv}. 
For the $\tau\to e\rho^0$ mode only we 
{also}
require that the magnitude of the 
thrust be in the range $0.90 < T < 0.96$
to reduce two-photon and inelastic $V^0$-production 
backgrounds.

\section{Signal and Background Estimation}

For all modes,
the $M_{\rm {\ell V^0}}$ and $\Delta E$  resolutions are 
{obtained}
from fits to the signal MC {distributions,}  
{using}  an asymmetric Gaussian function that takes into account 
{initial-state} radiation.
These {Gaussians} have widths
{as} shown in Table~\ref{tbl:width}.
\begin{table}
\begin{center}
\caption{Summary {of} $M_{\rm \ell V^0}$  
and
$\Delta E$
{resolutions} ($\sigma^{\rm{high/low}}_{M_{\rm{\ell V^0}}}$ (MeV/$c^2$) 
and 
$\sigma^{\rm{high/low}}_{\Delta E}$ (MeV)).
Here $\sigma^{\rm high}$ ($\sigma^{\rm low}$)
means the standard deviation
{on the} higher (lower) side of the peak.}
\label{tbl:width}
\vspace*{0.2cm}
\begin{tabular}{c|cccc} \hline\hline
Mode
& $\sigma^{\rm{high}}_{M_{\rm{\ell V^0}}}$
& $\sigma^{\rm{low}}_{M_{\rm{\ell V^0}}}$ 
& $\sigma^{\rm{high}}_{\Delta E}$ 
&  $\sigma^{\rm{low}}_{\Delta E}$ 
 \\ \hline
$\tau\to\mu\rho^0$
& 6.1  &  5.4 & 16.0 & 21.9 \\
$\tau\to e\rho^0$
& 6.7   &  5.7 & 15.6 &25.1 \\
$\tau\to\mu \phi$
& 3.7  &  3.8 & 14.2 & 19.9 \\
$\tau\to e\phi$
& 4.1  &  4.5 & 14.0 & 22.0 \\
$\tau\to\mu \omega$
& 7.0  &  8.9 & 25.7 & 29.0 \\
$\tau\to e\omega$
& 8.6  &  9.7 & 21.1 & 37.1 \\
$\tau\to\mu K^{\ast 0}$
& 4.9  &  5.2 & 15.8 & 21.2 \\
$\tau\to eK^{\ast 0}$
& 5.7  &  6.7 & 15.6 & 25.1 \\
$\tau\to\mu \bar{K}^{\ast0}$
& 4.9  &  5.2 & 15.8 & 21.3 \\
$\tau\to e\bar{K}^{\ast0}$
& 5.2  &  5.7 & 15.6 & 24.6 \\
\hline\hline
\end{tabular}
\end{center}
\end{table}

{To evaluate  branching fractions, we use elliptical signal 
regions {with} 
major and minor axes equal to  $3\sigma$ in
the $\Delta E$ and $M_{\ell V^0}$ distributions, where $\sigma = 
(\sigma^{\rm high}+\sigma^{\rm low})/2$ from Table~\ref{tbl:width}.
The center and rotation angle are determined by scanning to
maximize the efficiency of the signal divided by the area.} 
We blind the data in the signal region
until all selection criteria are finalized
so as not to bias our choice of selection criteria.

{Figures}~\ref{fig:muv0}  and \ref{fig:ev0} 
show scatter-plots 
for data events and signal MC samples 
distributed over $\pm 20\sigma$ 
in the $M_{\ell V^0}-\Delta E$ plane for the $\tau\to \mu V^0$ and $e V^0$ modes, 
respectively.
For the $\tau\to eV^0$ modes, {the dominant background comes} 
from two-photon processes,
while the fraction of $q\bar{q}$ and generic $\tau^+\tau^-$ events 
is small due to {the} low electron fake rate.
For the $\tau\to \mu \rho^0$ mode,
{the} dominant background 
{comes from}
$e^+e^-\to q\bar{q}$ events. 
{{A} 
smaller background of}
generic
$\tau^+\tau^-$ events in the {region}
$\Delta E<$ 0 GeV and 
$M_{\mu V^0} < $ $m_{\tau}$ {is due to}
combinations of a fake muon and two {pions.}
{For the $\tau\to \mu \omega$   mode,
{%a small  background around the signal region comes 
{there is a small background in the signal region}
from 
generic
$\tau^+\tau^-$ events
in which
{a} fake muon is combined with 
two pions from $\tau^-\to \pi^-\pi^+\pi^-\nu_{\tau}$ and 
a fake $\pi^0$ 
{(from ISR and beam background)}.
A large background 
{is} from $\tau^-\to h^- \omega \nu_{\tau}$ 
in the $\Delta E<$ 0 GeV and 
$M_{\mu V^0} < $ $m_{\tau}$ region.}
For the $\tau \to \mu \phi$ mode,
{the} 
dominant background 
{is due to}
$q\bar{q}$ 
and
$\tau^+\tau^-$ events 
{in which {a} pion is misidentified as a kaon.}
For the $\tau\to \mu K^{\ast0}$ and $\mu \bar{K}^{\ast0}$  modes,
the dominant background
{comes} 
from generic
$\tau^+\tau^-$ events
{with}
a fake muon, fake kaon  and real pion
from $\tau^-\to \pi^-\pi^+\pi^-\nu_{\tau}$.
If {one of the pions} 
is misidentified as a kaon,
{then} the reconstructed mass from
generic $\tau^+\tau^-$ background 
could be greater than the $\tau$ lepton mass, 
since the kaon mass is greater than {that} of the pion.

For the $\mu\rho^0$ and $\mu\omega$ {modes,}
{where significant background remains after applying 
%previous 
{the above-mentioned}
selection criteria,} 
we extrapolate to the signal region 
%extrapolation to the signal region 
%is performed 
by fitting 
to observed data in 
{the $\pm 5\sigma_{\Delta E}$ $M_{\mu V^0}$ data sideband}
using {the sum of} 
an exponential
{and} {a} first-order polynomial function 
{for} generic {$\tau^+\tau^-$} 
and continuum, respectively.
For  the $\tau\to eV^0$,  $\mu\phi$,  $\mu K^{\ast0}$ and $\mu\bar{K}^{\ast0}$ 
modes, 
%{the remaining background is small, and}
{the background after all the event 
selection requirements have been applied is small;}
extrapolation to the signal region assumes that
the background distribution  is
flat along the {$M_{\ell V^0}$ axis within 
$\pm5\sigma_{\Delta E}$.}
We estimate {the} expected number 
of the background events in the signal
region for each  mode
using  the number of 
{data events observed} 
in the sideband region
inside 
{the} horizontal lines {but}
%excluding 
{outside}
the signal region.
The signal efficiency and 
%the number of expected background events 
{background level}
for each mode 
are {given} in Table~\ref{tbl:eff}.
After estimating the background,  we 
open {the} {blinded} 
regions.
We observe one candidate event for 
each of the $\tau\to \mu\phi$, $\mu K^{\ast0}$ and $\mu\bar{K}^{\ast0}$ modes,
and no candidate events for other modes.
{In {each case} the number} 
of events observed in the signal {region}
{is} 
consistent with
{the} expected number of background events.

\begin{figure}
\begin{center}
       \resizebox{0.36\textwidth}{0.34\textwidth}{\includegraphics
        {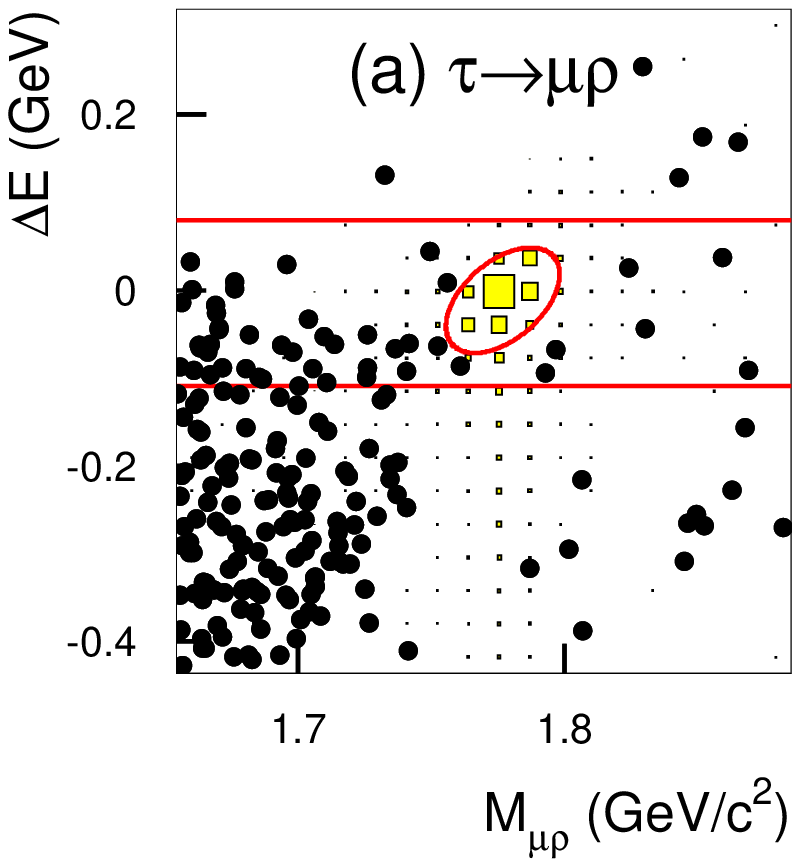}}
 \hspace*{-1.cm}
       \resizebox{0.36\textwidth}{0.34\textwidth}{\includegraphics
        {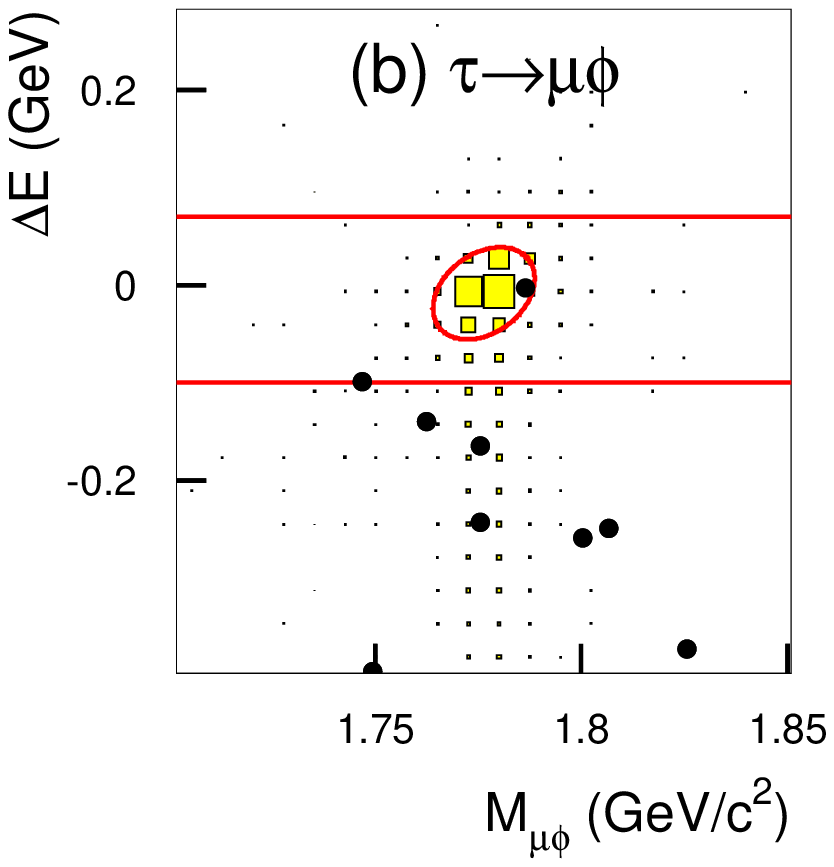}}
 \hspace*{-1.cm}
       \resizebox{0.36\textwidth}{0.34\textwidth}{\includegraphics
        {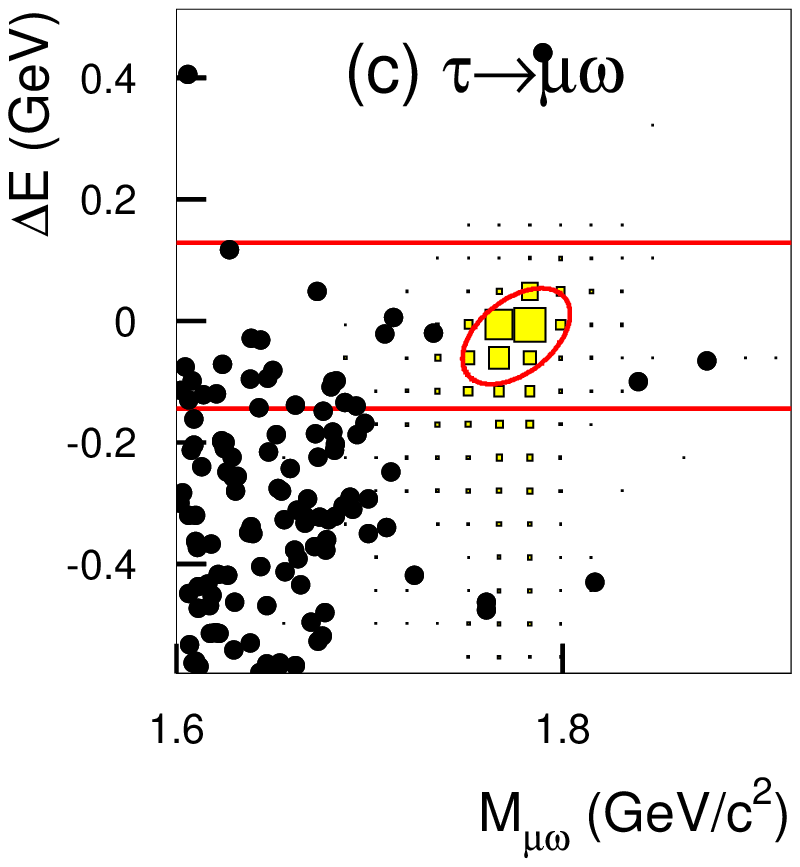}}\\
 \hspace*{-5.3cm}
       \resizebox{0.36\textwidth}{0.34\textwidth}{\includegraphics
        {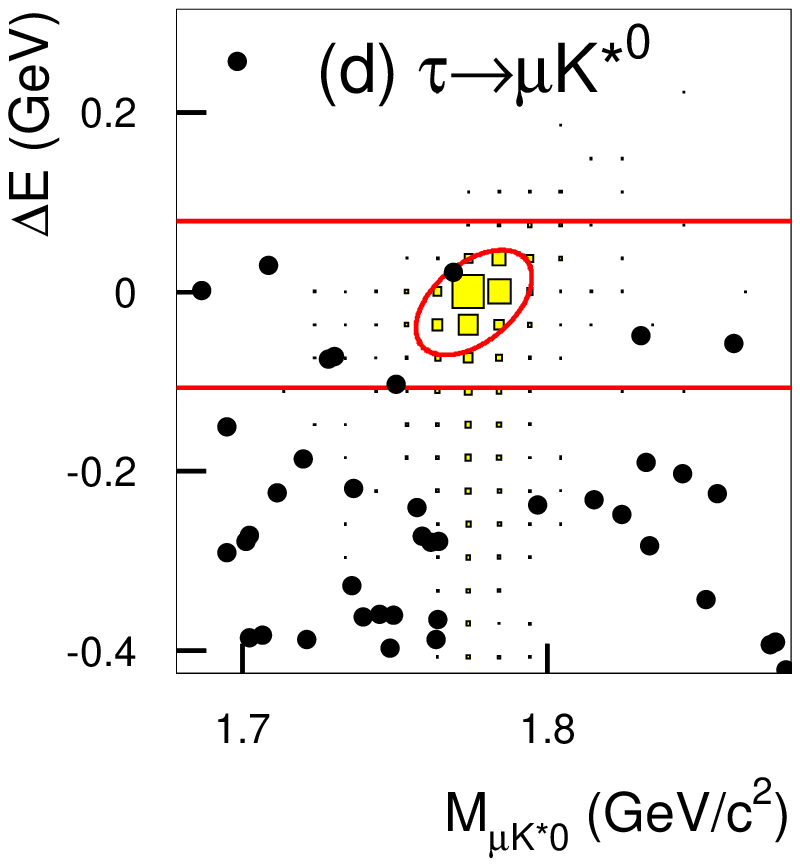}}
 \hspace*{-1.cm}
       \resizebox{0.36\textwidth}{0.34\textwidth}{\includegraphics
        {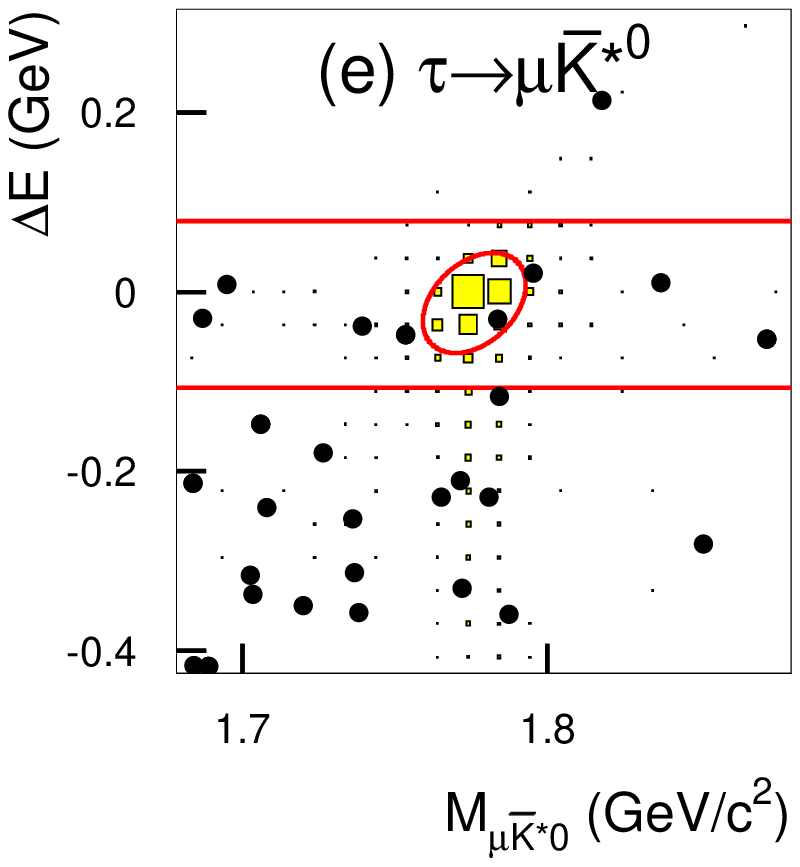}}
 \vspace*{-0.5cm}
\caption{Scatter-plots in 
{the} $\pm 20 \sigma$ area of  {the}
$M_{\mu V^0}$ -- $\Delta{E}$ plane:
(a), (b), (c), (d) and (e) show
the $\mu\rho^0$, $\mu\phi$, $\mu\omega$, $\mu {K}^{\ast0}$ and 
$\mu \bar{K}^{\ast0}$ modes, respectively. 
{Experimental data events are shown as}
solid circles.
The filled boxes show the MC signal distributions
with arbitrary normalization.
The elliptical signal (blind)
{regions} 
shown by  solid curves 
are used for evaluating the signal yield.
The region between the horizontal solid lines 
(excluding the signal region) is
used to estimate the expected background in the elliptical region. 
}
\label{fig:muv0}
\end{center}
\end{figure}

\begin{figure}
\begin{center}
       \resizebox{0.36\textwidth}{0.34\textwidth}{\includegraphics
        {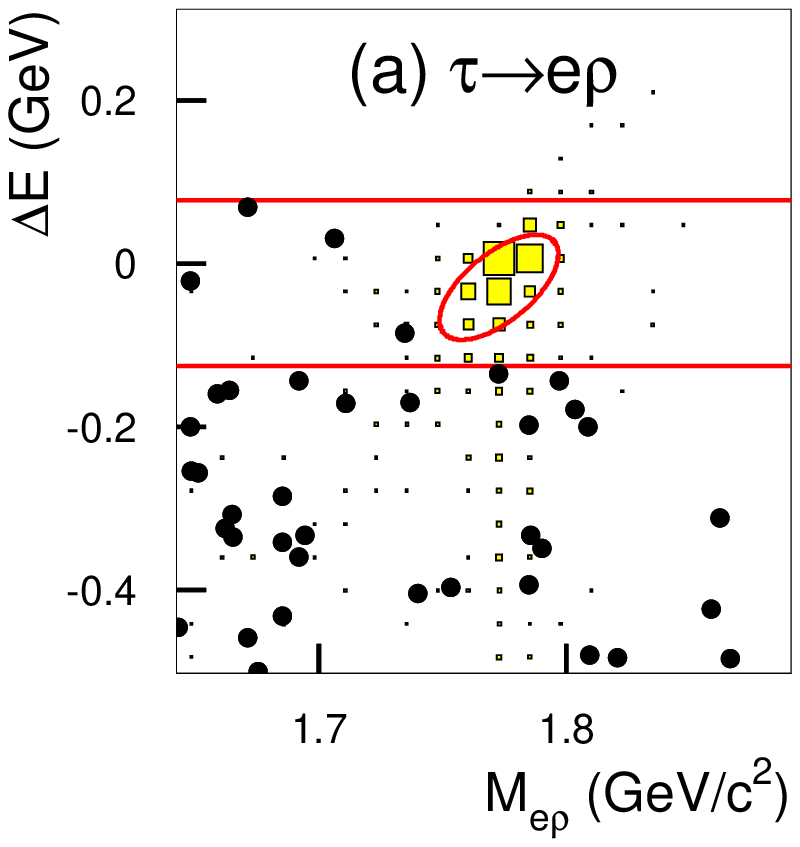}}
 \hspace*{-1.cm}
       \resizebox{0.36\textwidth}{0.34\textwidth}{\includegraphics
        {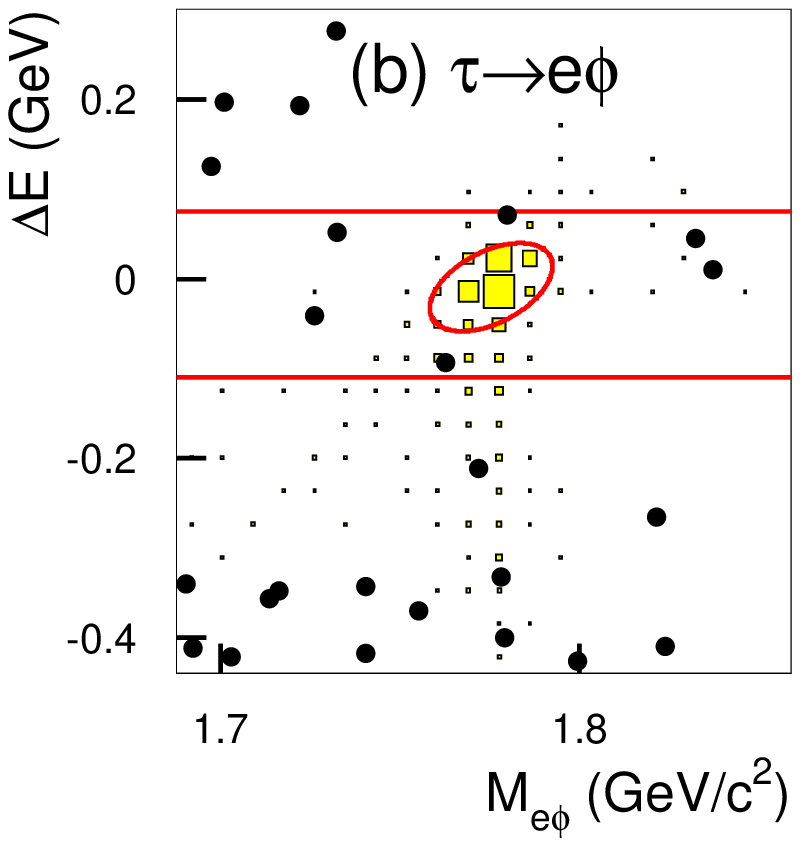}}
 \hspace*{-1.cm}
       \resizebox{0.36\textwidth}{0.34\textwidth}{\includegraphics
        {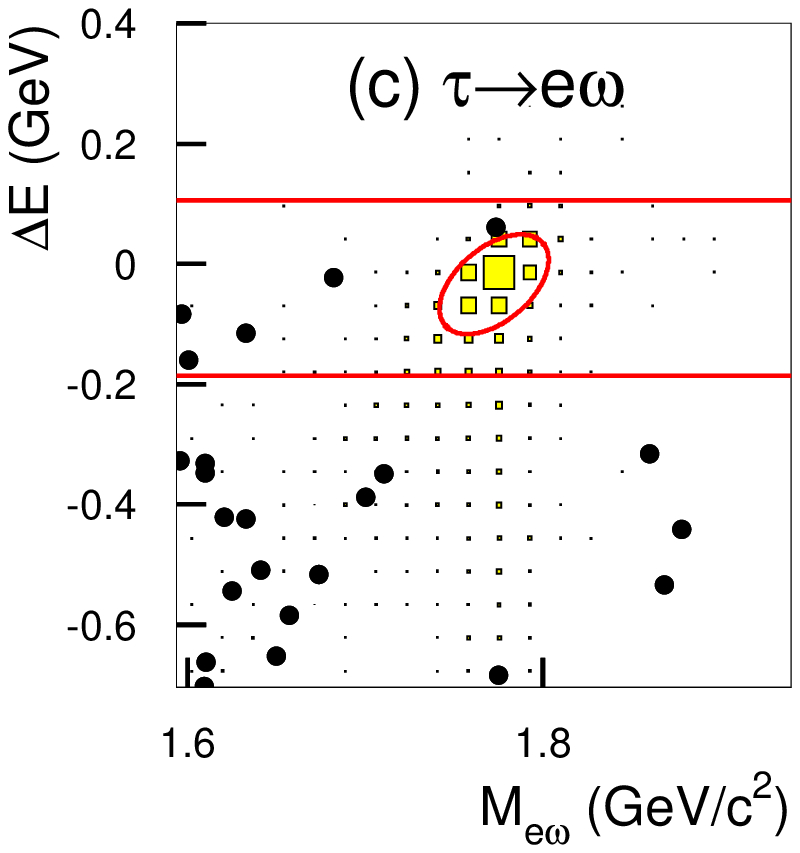}}\\
 \hspace*{-5.3cm}
       \resizebox{0.36\textwidth}{0.34\textwidth}{\includegraphics
        {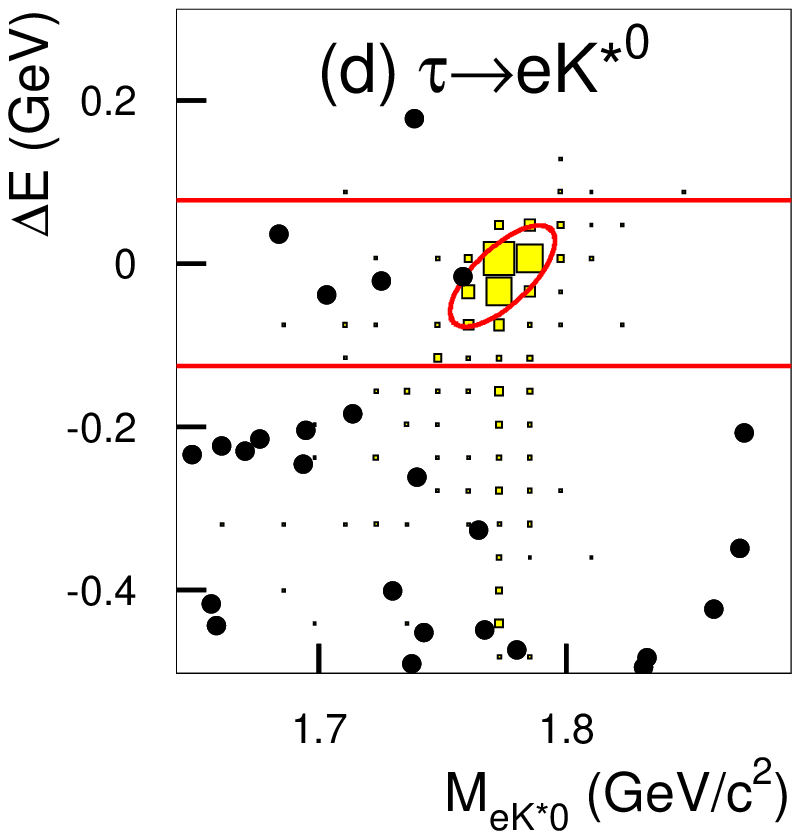}}
 \hspace*{-1.cm}
       \resizebox{0.36\textwidth}{0.34\textwidth}{\includegraphics
        {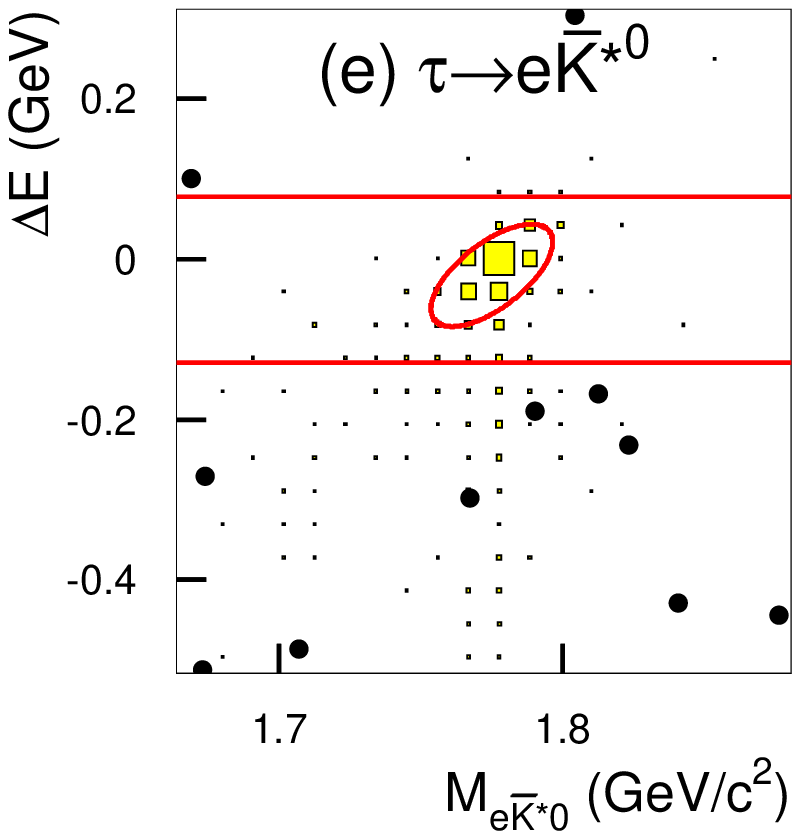}}
 \vspace*{-0.5cm}
\caption{Scatter-plots in the
$\pm 20 \sigma$ area of {the}
$M_{eV^0}$ -- $\Delta{E}$ plane: 
(a), (b), (c), (d) and (e) show
the $e\rho^0$, $e\phi$, $e\omega$, $e {K}^{\ast0}$ and 
$e \bar{K}^{\ast0}$ modes, respectively. 
{Experimental data events are shown as}
solid circles.
The filled boxes show the MC signal distributions
with arbitrary normalization.
The elliptical signal (blind)
{regions} 
shown by  solid curves 
are used for evaluating the signal yield.
The region between the horizontal solid lines (excluding
the signal region) is
used to estimate the expected background in the elliptical region. 
}
\label{fig:ev0}
\end{center}
\end{figure}

The dominant systematic uncertainties
{in} this analysis
{relate to}
tracking efficiencies
and particle identification.
The uncertainty due to the {track finding efficiency} 
is  estimated to be 1.0\% per charged track,
therefore, the total uncertainty due to the charged track finding is
4.0\%.
{The uncertainty due to particle {identification} 
{is} 
2.2\%, 1.9\%, 
1.3\% and 1.8\% per electron, muon, pion and kaon, respectively.}
The uncertainties due to the {$V^0$ branching {fractions} 
are} estimated to be {1.2\%} 
and 0.7\% 
for the $\phi$ and $\omega$, respectively.
The {uncertainty due to {the} integrated} 
luminosity is estimated to be {1.4\%.}
The uncertainties due to the trigger efficiency and MC statistics  
are negligible 
compared with the other uncertainties.
All these uncertainties are added in quadrature, 
and the total systematic uncertainties for  
{the different modes range from 5.3\% to 6.2\%,
as shown in Table~\ref{tbl:eff}.}

\begin{table}
\begin{center}
\caption{ The signal {efficiency ($\varepsilon$),} 
the number of expected background {events}  ($N_{\rm BG}$)
estimated from the  sideband data, 
{total} 
systematic uncertainty  ($\sigma_{\rm syst}$),
{the} number of observed events 
in the signal region ($N_{\rm obs}$), 
90\% C.L. upper limit on the number of signal events including 
systematic uncertainties~($s_{90}$), 
90\% C.L. upper limit on 
{the}
observed branching  fraction  (${\cal{B}}_{\rm obs.}$)
for each individual mode. }
\label{tbl:eff}
\begin{tabular}{c|cccccc}\hline \hline
Mode &  $\varepsilon$~{(\%)} & 
$N_{\rm BG}$  & $\sigma_{\rm syst}$~{(\%)}
& $N_{\rm obs}$ 
& ${s}_{90}$  
&~~${\cal{B}}_{\rm obs}$ $(\times10^{-8})$  
\\ \hline
$\tau^-\to \mu^-\rho^0 $ & 7.09 & $1.48\pm{0.35}$ &5.3 &
 0 & 1.34 & 1.2 
\\

$\tau^-\to e^-\rho^0 $ & 7.58 & $0.29\pm{0.15}$ & 5.4 &
  0  & 2.17 & 1.8 
\\

 $\tau^-\to \mu^-\phi $ &  3.21& $0.06\pm{0.06}$ & 5.8 &
 1 &  4.24&8.4 
\\
$\tau^-\to e^-\phi $ &4.18 & $0.47\pm{0.19}$ & 5.9 &
 0   & 2.02 &3.1
\\

$\tau^-\to \mu^-\omega $ & 2.38 & $0.72\pm{0.18}$ & 6.1 &
 0   & 1.76 &4.7 
\\
$\tau^-\to e^-\omega $ & 2.92 & $0.30\pm{0.14}$ & 6.2 &
 0   & 2.19 &4.8 
\\

$\tau^-\to \mu^-K^{\ast0} $ & 3.39 & $0.53\pm{0.20}$ & 5.5 &
 1   & 3.81 &7.2 
\\
$\tau^-\to e^-K^{\ast0} $ & 4.37 & $0.29\pm{0.14}$ & 5.6 &
 0   & 2.17 &3.2 
\\

$\tau^-\to \mu^-\bar{K}^{\ast0} $ & 3.60 & $0.45\pm{0.17}$ & 5.5 &
 1   & 3.90 &7.0 
\\
$\tau^-\to e^-\bar{K}^{\ast0} $ & 4.41 & $0.08\pm{0.08}$ & 5.6 &
 0   & 2.34 & 3.4 
\\

\hline\hline
\end{tabular}
\end{center}
\end{table}

\section{Upper Limits on the branching fractions}

Since no statistically significant excess of data over
the expected background 
{{is} observed in the signal region,}
we set  upper limits on the branching fractions 
based on 
%the Feldman-Cousins method~\cite{cite:FC}.
{a frequentist approach~\cite{cite:FC}.}
{We calculate the} 
90\% C.L. upper limit on the number of signal events 
including  systematic uncertainty~($s_{90}$) 
{from} 
the number of expected {background events}, 
{the number of observed events}
and the systematic uncertainty,
{{using} the POLE program without conditioning~\cite{pole}.}
The upper limit on the branching fraction ($\cal{B}$) is then given by
\begin{equation}
{{\cal{B}}(\tau\to\ell V^0) <
\displaystyle{\frac{s_{90}}{2N_{\tau\tau}\varepsilon{}}}},
\end{equation}
where $N_{\tau\tau}$ is the number of $\tau^+\tau^-$pairs, and 
$\varepsilon$ is the signal efficiency including 
the branching fraction of $V^0$.
{The value {$N_{\tau\tau} =  782\times 10^6$}} is obtained 
from 
{the} integrated luminosity times 
the cross section of {$\tau$-pair production,} which 
is calculated 
{in 
KKMC~\cite{tautaucs} to be 
$\sigma_{\tau\tau} = 0.919 \pm 0.003$ nb and
$\sigma_{\tau\tau} = 0.875 \pm 0.003$ nb 
for 782 fb$^{-1}$ {at} $\Upsilon(4S)$ 
and 72 fb$^{-1}$ {at} $\Upsilon(5S)$, respectively.}
Table~\ref{tbl:eff}
summarizes
information about 
upper limits 
for all modes.
We obtain the following ranges for 90\% C.L. upper limits 
{on the branching fractions:} 
${\cal{B}}(\tau\rightarrow eV^0)
 < (1.8-4.8)\times 10^{-8}$
and 
${\cal{B}}(\tau\rightarrow \mu V^0)< (1.2-8.4)\times 10^{-8}$.
These results improve {upon} 
our previously published upper limits~\cite{lv0_belle}
by factors {of} up to 5.7.
They are also more 
{restrictive}
than 
the recent results 
from BaBar~\cite{lv0_babar, lomega_babar}.

\section{Summary}
We have searched for {lepton-flavor-violating} $\tau$ decays 
into a lepton and a vector meson
using 854 fb$^{-1}$ of data
{collected} 
with the Belle detector at the 
KEKB asymmetric-energy $e^+e^-$ collider.
Since no evidence for a signal 
is {found, we set the} following 90\% C.L. upper limits 
{on the branching fractions:} 
${\cal{B}}(\tau\rightarrow eV^0)
 < (1.8-4.8)\times 10^{-8}$
and 
${\cal{B}}(\tau\rightarrow \mu V^0)< (1.2-8.4)\times 10^{-8}$.
{These results improve {upon} 
our previously published upper limits
by factors {of} up to 5.7.
This improvement results both from 
using {a} 
larger data sample and from
 an improved rejection
of specific backgrounds, 
such as di-baryon production in the continuum 
for the $\tau\to\mu V^0$ modes, 
and $\tau^-\to h^-\pi^0\nu_{\tau}$ decays with a photon conversion
for the $\tau\to eV^0$ modes.} 
{The} more stringent upper limits {reported here} 
can be used to constrain the {parameter space} in various models of new physics.

\section*{Acknowledgments}

%% Please paste this acknowledgement into your latex file.
% updated  4/02/10   WCU (Korea) added in short version
% updated  2/15/10   Czech new contract no. added
% updated  1/06/10   Korean section updated, Czech section added
% updated  4/26/09   Korean section updated (add KISTI)
%
%***** Acknowledgments *****
%----------- Long version, for most papers ----------- 
We thank the KEKB group for the excellent operation of the
accelerator, the KEK cryogenics group for the efficient
operation of the solenoid, and the KEK computer group and
the National Institute of Informatics for valuable computing
and SINET3 network support.  We acknowledge support from
the Ministry of Education, Culture, Sports, Science, and
Technology (MEXT) of Japan, the Japan Society for the 
Promotion of Science (JSPS), and the Tau-Lepton Physics 
Research Center of Nagoya University; 
the Australian Research Council and the Australian 
Department of Industry, Innovation, Science and Research;
the National Natural Science Foundation of China under
contract No.~10575109, 10775142, 10875115 and 10825524; 
the Ministry of Education, Youth and Sports of the Czech 
Republic under contract No.~LA10033 and MSM0021620859;
the Department of Science and Technology of India; 
the BK21 and WCU program of the Ministry Education Science and
Technology, National Research Foundation of Korea,
and NSDC of the Korea Institute of Science and Technology Information;
the Polish Ministry of Science and Higher Education;
the Ministry of Education and Science of the Russian
Federation and the Russian Federal Agency for Atomic Energy;
the Slovenian Research Agency;  the Swiss
National Science Foundation; the National Science Council
and the Ministry of Education of Taiwan; and the U.S.\
Department of Energy.
This work is supported by a Grant-in-Aid from MEXT for 
Science Research in a Priority Area (``New Development of 
Flavor Physics''), and from JSPS for Creative Scientific 
Research (``Evolution of Tau-lepton Physics'').


\begin{thebibliography}{99}


\bibitem{rpv}
J.~E.~Kim, P.~Ko and D.~G.~Lee,
Phys. Rev. D {\bf 56}, 100 (1997).


\bibitem{cite:amon}
A.~Ilakovac, Phys. Rev. D {\bf 62}, 036010 (2000).

\bibitem{cite:six_fremionic}
D.~Black {\it et al.},
Phys.\ Rev.  D {\bf  66}, 053002 (2002).

\bibitem{cite:susy1}
        C.-H.~Chen and C.-Q.~Geng,
        Phys.~Rev.~D {\bf 74}, 035010 (2006).

\bibitem{cite:susy2}
 E.~Arganda {\it et al.}, JHEP {\bf 0806}, 079 (2008).


\bibitem{Benbrik:2008ik}
  R.~Benbrik and C.~H.~Chen,
  %``Leptoquark on $P\to \ell^{+} \nu$, FCNC and LFV,''
  Phys.\ Lett.\  B {\bf 672}, 172 (2009).

\bibitem{Li:2009yr}
  Z.~H.~Li, Y.~Li and H.~X.~Xu,
  %``Unparticle-Induced Lepton Flavor Violating Decays \tau^- ->l^- (V^0,
  %~P^0),''
  Phys.\ Lett.\  B {\bf 677}, 150 (2009).


\bibitem{Belle}
A.~Abashian {\it et al.} (Belle Collaboration),
Nucl. Instr. and Meth. A {\bf 479}, 117 (2002).

\bibitem{kekb}
S.~Kurokawa and  E.~Kikutani, Nucl. Instr. {and} Meth. A {\bf 499}, 1
(2003), and other papers included in this Volume.


\bibitem{lv0_belle}
  {Y.~Nishio {\it et al.}  (Belle Collaboration),
  Phys.\ Lett.\  B {\bf 664}, 35 (2008).}

\bibitem{lv0_babar}
  B.~Aubert {\it et al.}  (BaBar Collaboration),
  Phys.\ Rev.\ Lett.\  {\bf 103}, 021801 (2009).

\bibitem{lomega_babar}
  B.~Aubert {\it et al.}  (BaBar Collaboration),
  Phys.\ Rev.\ Lett.\  {\bf 100}, 071802 (2008).






\bibitem{EID}
        K.~Hanagaki {\it et al.},
        Nucl. Instr. and Meth.  A {\bf 485}, 490 (2002).



\bibitem{MUID}
	{A.~Abashian {\it et al.}},
        Nucl. Instr. and Meth. A 
	{\bf 491}, 69 (2002).


\bibitem{KKMC}
        S.~Jadach {\it et al.},
        Comp. Phys. Commun. {\bf 130}, 260 (2000).


\bibitem{evtgen}
	D.~J.~Lange,
        {Nucl. Instr. and Meth. A} 
	{\bf 462}, 152 (2001).




\bibitem{BHLUMI}
        S.~Jadach {\it et al.},
        Comp. Phys. Commun. {\bf 70}, 305 (1992).


\bibitem{AAFH}
        F.~A.~Berends {\it et al.},
        Comp. Phys. Commun. {\bf 40}, 285 (1986).




\bibitem{thrust}
S.~Brandt {\it et al.},
        Phys.\ Lett.\ {\bf 12}, 57 (1964);
E.~Farhi,
        Phys.\ Rev.\ Lett.\ {\bf 39}, 1587 (1977).

\bibitem{cite:tau_egamma}
K.~Hayasaka  {\it et al.} (Belle Collaboration),
Phys.\ Lett.\ B {\bf  613}, 20 (2005).


\bibitem{cite:FC}
        G.~J.~Feldman and {R.~D. Cousins,}
        Phys.\ Rev.\ D {\bf 57}, 3873 (1998).



\bibitem{pole}
       See {http://www3.tsl.uu.se/${}^{\sim}$conrad/pole.html;}
        J.~Conrad {\it et al.},
        Phys.\ Rev.\ D {\bf 67}, 012002 (2003).    


\bibitem{tautaucs}
        S.~Banerjee {\it et al.},
        Phys.\ Rev.\ D {\bf 77}, 054012 (2008).

%\bibitem{PDG}
%W.-M.~Yao   {\it et al.} (Particle Data Group),  
%	{J. Phys. G} {\bf  33}, 1 (2006).


\end{thebibliography}
\end{document}